\documentclass{PoS}
\usepackage{color}

\title{Particle Acceleration in Astrophysical Sources}

\ShortTitle{Particle Acceleration in Astrophysics}

\author{\speaker{Elena Amato}\thanks{A footnote may follow.}\\
        INAF - Osservatorio Astrofisico di Arcetri, Largo E. Fermi, 5, I-50125, Firenze, Italy\\
        E-mail: \email{amato@arcetri.astro.it}}

\abstract{Astrophysical sources are extremely efficient accelerators. Some sources emit photons up to multi-TeV energies, a signature of the presence, within them, of particles with energies much higher than those achievable with the largest accelerators on Earth. Even more compelling evidence comes from the study of Cosmic Rays, charged relativistic particles that reach the Earth with incredibly high energies: at the highest energy end of their spectrum, these subatomic particles are carrying a macroscopic energy, up to a few Joules. 

Here I will address the best candidate sources and mechanisms as cosmic particle accelerators. I will mainly focus on Galactic sources such as Supernova Remnants and Pulsar Wind Nebulae, which being close and bright, are the best studied among astrophysical accelerators. These sources are held responsible for most of the energy that is put in relativistic particles in the Universe, but they are not thought to accelerate particles up to the highest individual energies, $\approx 10^{20}$ eV. However they allow us to study in great detail acceleration mechanisms such as shock acceleration (both in the newtonian and relativistic regime) or magnetic reconnection, the same processes that are likely to be operating also in more powerful sources.}

\FullConference{Frontiers of Fundamental Physics 14 - FFP14,\\
                15-18 July 2014\\
                Aix Marseille University (AMU) Saint-Charles Campus, Marseille }
		
\begin{document}

\section{Introduction}
\label{sec:intro}
The flux of Cosmic Rays (CRs) impinging on Earth provides striking evidence of the existence of very effective accelerators in the Cosmos. As we have known for over a century, CRs are charged energetic particles: mainly protons, electrons and nuclei, but also traces of anti-matter. At low energies, they mostly come from within the solar system, while with increasing energy they reach us from distant sources, within our Galaxy and beyond. Their energy spectrum extends over at least 13 decades in energy and the most energetic particles detected have energies of a few Joules, about 10 million times larger than the maximum energy achievable at the LHC.

\begin{figure}[h!!!]
\includegraphics[scale=.28]{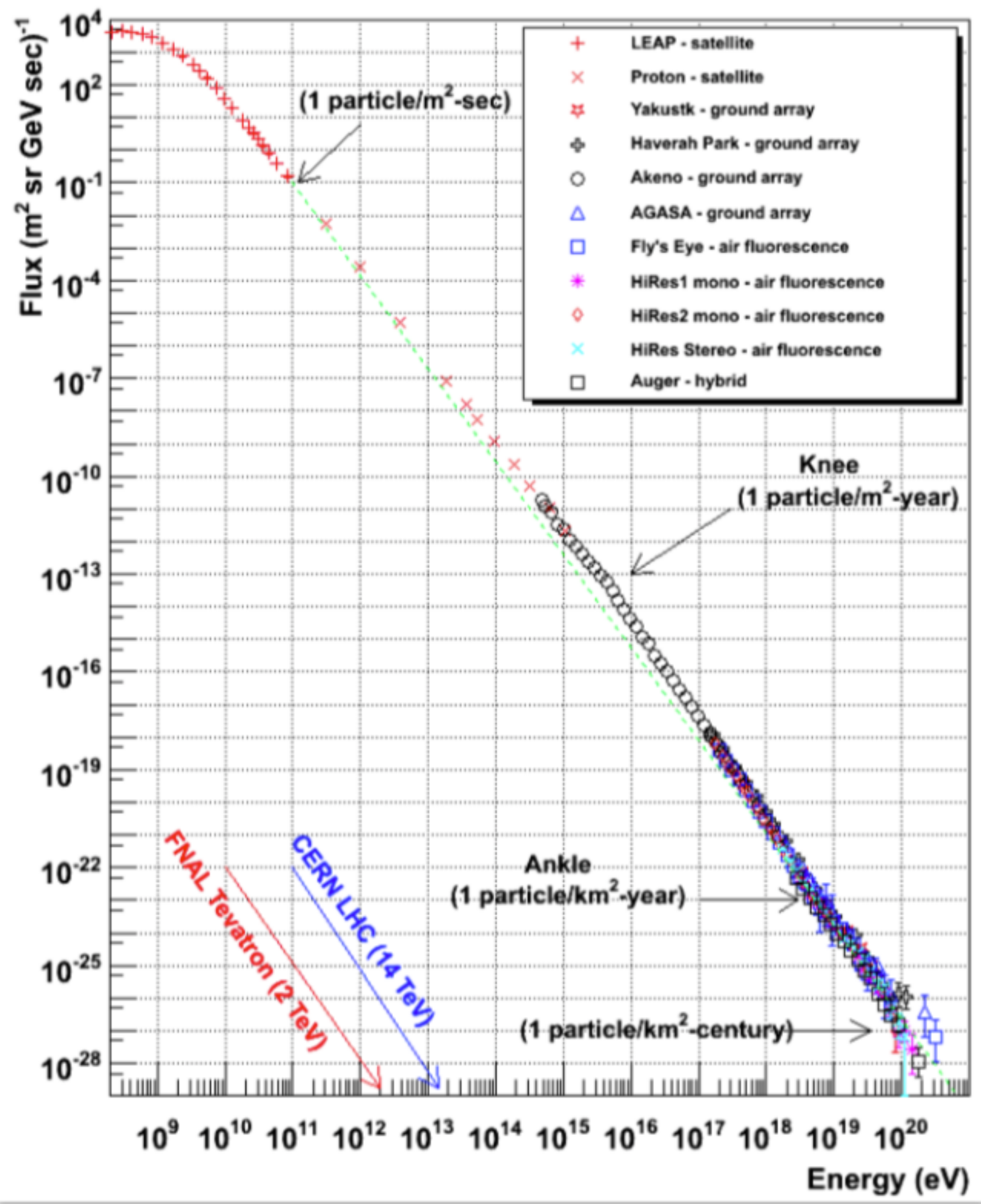}
\includegraphics[scale=.27]{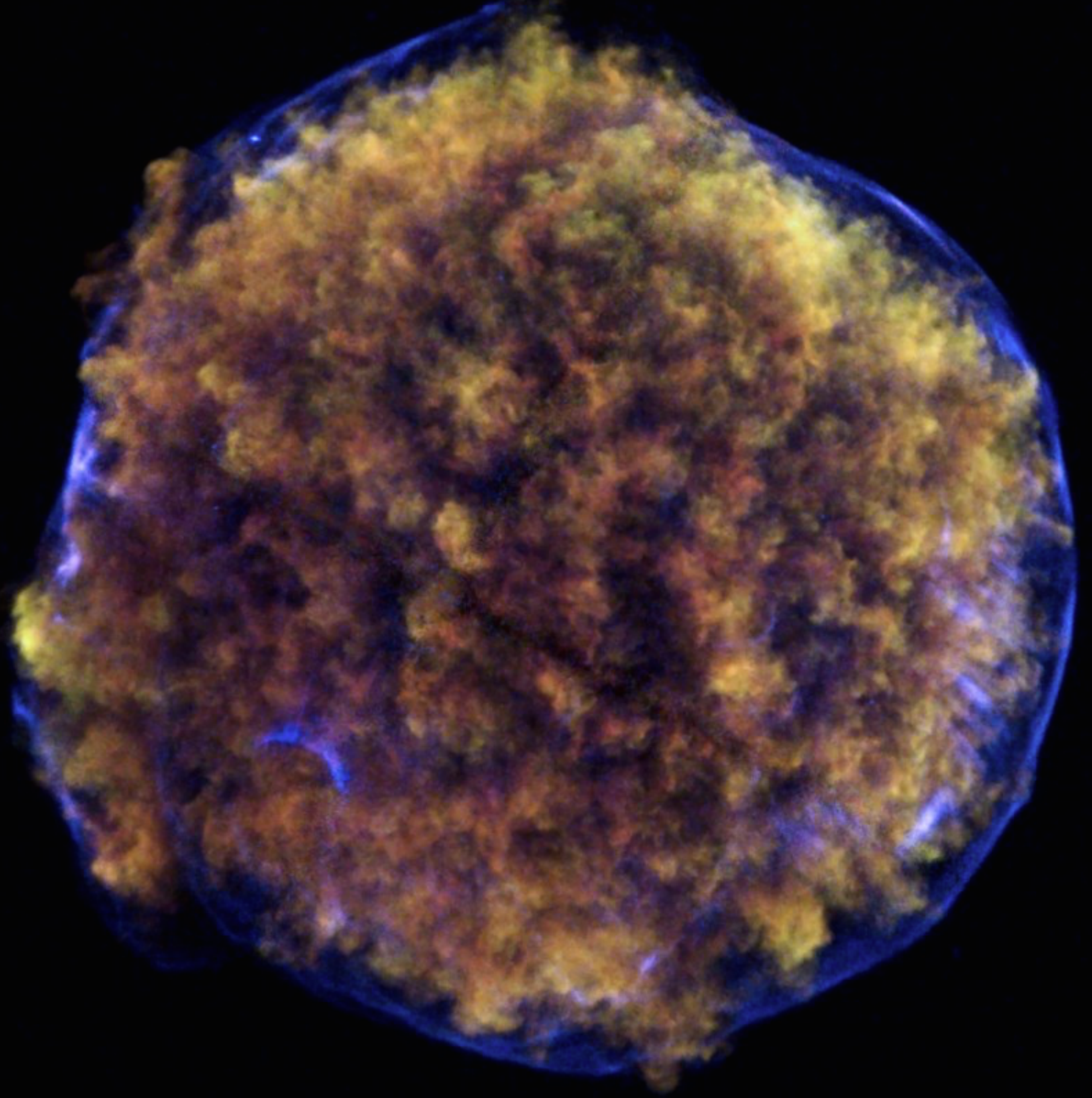}
\caption{Left panel: measured spectrum of Cosmic Rays from Ref.~\cite{blasirev}. Right panel: the Tycho SNR seen in X-rays (Image credit: NASA / CXC / F.J. Lu et al.)}
\label{fig:CRspec}
\end{figure}

Over this very wide range of energies the all-particle spectrum, obtained as the sum of all CRs regardless of their species, is very smooth (see left panel of Fig.~\ref{fig:CRspec}), and well described as a sequence of few power-laws, starting from energies $\approx$ 30 GeV (above which the effects of the solar wind on particle transport cease to be important). The most evident changes of spectral slope occur at the so-called {\it knee} at $E_{\rm knee}\approx 3 \times 10^{15}$eV, where the power-law index steepens from 2.7 to 3.1, and at the so-called {\it ankle} at about $10^{18}$ eV, where a flattening is observed instead. The first change of slope is usually taken to signal the maximum energy up to which protons are accelerated in the Galaxy. The general consensus is that the acceleration mechanism is rigidity dependent and nuclei with charge Z can be accelerated to Z times higher energies. This places the transition between galactic and extragalactic origin of CRs at energies between $10^{17}$ and $10^{19}$ eV.

The lowest energy CRs, up to GeV energies, come from the Sun and the solar wind, while particles with energies in the range between $\approx GeV$ and $\approx Z\ PeV$ are thought to be accelerated in Supernova Remnants (SNRs). At even larger energies, there is no general consensus on what the main sources are, but the most commonly invoked ones include Active Galactic Nuclei and their jets, Gamma Ray Bursts, newly born fast spinning pulsars and galaxy merger shocks \cite{blasiUHECR}.

While at the highest energy end of the spectrum the individual particle energies become really impressive, given the steepness of the spectrum, most of the total energy budget in CRs resides in low energy galactic particles. And in spite of their very small number density, $n_{\rm CR}\approx 10^{-9} {\rm cm}^{-3}$, the energy density these particles embed is large, comparable with that of gas and magnetic field in the Galaxy. In the following I will mostly focus on the acceleration of these galactic particles, trying to summarise how well we understand what the main galactic accelerators are and how they work.

\section{The sources of Galactic Cosmic Rays}
\label{sec:GALCRS}

After being accelerated and leaving their sources, CR particles propagate diffusively in the galactic magnetic field. This makes it impossible to trace them back to their sources, and different types of arguments must be used to identify the latter. The SNR paradigm for the origin of galactic CRs is based on a very simple energetic argument: the luminosity that is needed to ensure the measured CR flux is $L_{\rm CR}\approx 3 \times 10^{40}$ erg/s (see \cite{amatorev} and \cite{blasirev} and references therein); a Supernova (SN) explosion typically leaves an expanding remnant with a kinetic energy $E_{\rm SN}\approx 10^{51}$erg. With the estimated frequency of SN explosions in our Galaxy, ${\cal R}\approx 1/(100 {\rm yr})$, $L_{\rm CR}$ corresponds to $\approx 10\%$ efficiency of conversion of $E_{\rm SN}$ into CRs. 

This association between CRs and SNRs has been around now for 80 years, but only in the last decade or so it has been possible to collect observational evidence in its favour. First direct proof that particle acceleration indeed takes place in SNRs came from the detection of non-thermal (synchrotron) X-ray emission from several young SNRs (see \cite{vink12} for a comprehensive review), showing that young SNRs are able to provide electron acceleration to multi-TeV energies. 

However CRs are mostly protons and a lot of efforts were put in the quest for evidence of hadronic acceleration. The main channel through which protons radiate is $\gamma$-rays due to the decay of neutral pions produced by nuclear collisions of relativistic protons with the interstellar gas. The TeV detection of a few young SNRs appeared very promising in this respect, but unambiguous identification of hadronic $\gamma$-rays turned out to be impossible: TeV photons can come from $\approx 100$ TeV protons undergoing nuclear collisions, but also from TeV electrons upscattering the microwave and infrared background (Inverse Compton scattering process), and in most cases the data do not allow one to disentangle the two contributions. 

Only very recently direct evidence of hadronic acceleration in SNRs has finally also come, thanks to the lower energy $\gamma$ ray observatories {\it AGILE} and {\it Fermi}. Observing middle-aged SNRs (a few tens of thousands year old) interacting with molecular clouds (where the target density for nuclear collisions is largely enhanced, being the gas density much higher than in the diffuse ISM), these instruments have detected $\gamma$ ray emission with a spectrum unmistakably associated with $\pi^0$ decay at least from two sources, W44 \cite{w44fermi,w44agile} and IC443 \cite{ic443agile,ic443fermi}.

While this piece of evidence finally showing relativistic hadrons associated with SNRs is extremely important, it cannot be considered as conclusive in terms of proving the SNR paradigm. If SNRs are the main sources of all Galactic CRs they must be able to accelerate protons up to PeV energies, whereas in W44 and IC443 the proton spectrum is cut off at a few $\times 100$ GeV. This is not too surprising given the age of the sources: PeV particles are thought to be accelerated at the high speed shock of young sources. Exactly how young is a question on which there have been recent important developments to be discussed later on.

Also young sources when observed with {\it Fermi}'s eyes were found to be surprising, showing features opposite to what the basic acceleration theory was predicting. We will discuss these findings after a brief summary of the theory.

\section{Particle acceleration in SNRs}
\label{sec:nldsa}
The most commonly invoked particle acceleration mechanism in Astrophysics is diffusive shock acceleration (DSA), also known as $1^{\rm st}$ order Fermi process. The idea behind this process is that the particle gains energy each time it crosses the shock from upstream to downstream and vice versa, and this happens because at each crossing the particle always suffers a head on collision with the magnetic irregularities in the fluid on the other side of the shock. The energy gain at each crossing is $(\Delta E/E) \propto (V_s/c)$, where $V_s$ is the shock velocity.

A very attractive feature of DSA is the fact that at a strong shock it gives rise to a particle spectrum which is a power-law in momentum with a universal slope, close to what is implied from CR observations \cite{bell78}. In general the spectrum of shock accelerated particles will be given by $N(p)\propto p^{-\gamma_p}$, where $N(p)$ is the number of particles per unit momentum interval $dp$ and $\gamma_p=3 R_T/(R_T-1)$ with $R_T$ the compression ratio at the shock, namely $R_T=u_1/u_2$ with $u_1$ and $u_2$ the fluid velocities upstream and downstream of the shock respectively. The compression factor $R_T$ only depends on the shock Mach number, $M_S=u_1/C_{s1}$, where $C_{s1} \approx 10 \sqrt{T_4}$ km/s is the sound speed in ISM, whose temperature $T$ has been expressed in units of $10^4$ K. Using the standard Rankine-Hugoniot relations to describe the jump of all thermodynamical quantities at the shock, one finds $R_T=4 M_s^2/(3+M_s^2)$. For strong shocks, $M_s\gg1$, the latter expression reduces to $R_T=4$ and hence $\gamma_p=4$. For relativistic particles ($E\gg mc^2$), this slope in momentum is equivalent to a slope in energy that is easily calculated using the fact that $N(E) dE=E^{-\gamma_e} dE=4 \pi p^2 N(p)dp$. The result is $\gamma_e=\gamma_p-2$ which for strong shocks gives $\gamma_e=2$. This is exactly what is required to explain the CR spectrum at energies below the knee, if propagation effects (to be discussed later) lead to a steepening by $\sim 0.7$. The injection spectral index will be $\gamma_e>2$ for weaker shocks.

A noticeable feature of this process is that the particle spectrum is totally insensitive to the scattering properties. This is because the probability for particles to return to the shock is unaffected by scattering. What does depend on scattering, however, is the time it takes for the particles to get back to the shock, and hence the maximum number of crossings a particle can undergo during the lifetime of the system, or before being affected by energy losses: in other words, the maximum achievable energy, $E_{\rm max}$. In the diffusive regime, the time it takes for a particle to complete a cycle around the shock is:
\begin{equation}
t_{\rm acc}=\frac{3}{u_1-u_2}\left[\frac{D_1}{u_1}+\frac{D_2}{u_2}\right]
\label{eq:tacc}
\end{equation}
where $D_1$ ($u_1$) and $D_2$ ($u_2$) are the diffusion coefficients (fluid velocity) upstream and downstream of the shock, which depend on the particle energy and on the level of magnetic turbulence. $E_{\rm max}$ is then determined by the condition that the acceleration time be less than the age of the system and the timescale for losses: $t_{\rm acc}(E_{\rm Max})=\min(t_{\rm age}, t_{\rm loss})$. While losses are usually not a concern for hadrons, $t_{\rm age}$ tends to limit the maximum achievable energy well below the {\it knee} for standard values of the SNR parameters. Indeed one usually identifies $t_{\rm age}$ with the Sedov time $T_{\rm Sedov}$, when the expansion of the SNR starts decelerating: this happens when the blast wave has swept up a mass equal to that ejected by the SN event. For typical SN parameters one finds $T_{\rm Sedov}\approx 200 {\rm yr}$ for a SNR that is expanding in the normal ISM (density of $\approx 1 {\rm cm}^{-3}$).

The other missing ingredient to estimate the maximum achievable energy at a SNR shock is the diffusion coefficient. The motion of particles is determined by their interactions with magnetic perturbations. In the case of low frequency waves, these lead to pitch angle scattering (change in the direction of particle motion) and hence spatial diffusion. The diffusion coefficient can be written, in quasi-linear theory, as \cite{blandeich}:
\begin{equation}
D(p)=\frac{4}{3\pi}\left(\frac{B_0}{\delta B}\right)^2 c\ r_L\ ,
\label{eq:diffcoeff}
\end{equation}
where $B_0$ is the large scale magnetic field, $r_L=cp/(eB_0)$ the Larmor radius of a particle of momentum $p$ in the unperturbed magnetic field and $\delta B$ is the perturbation at a wavelength resonant with the particle's orbit $\lambda \approx r_L$. The energy dependence of the diffusion coefficient is then determined by the spectrum of the magnetic perturbations through the resonance condition.
The most common assumptions are that of a Kolmogorov spectrum ${\cal F} (k)\propto k^{-5/3}$ leading to $\delta B(k)^2\propto k^{-2/3}$ and hence $D(p)\propto p^{1/3}$, and that of Bohm diffusion, assuming constant power at all scales ${\cal F} (k)\propto k^{-1}$ and $\delta B\approx B_0$, leading to $D(p)\approx D_B(p)= c r_L/3$. 
The Kolmogorov spectrum is appropriate to describe a hydrodynamic cascade, and has long been considered as a good description of the turbulence spectrum in the Galaxy, where turbulence is thought to be primarily seeded by SN explosions with $\delta B/B\approx 1$ at an injection scale of 50-100 pc. If one assumes that such a description of the turbulence is what determines the diffusion of particles also at a SN shock, the maximum energy that would be computed from Eq.~\ref{eq:tacc} is ridiculously small: $E_{\rm max}\approx 1 GeV$ and even if one assumes that turbulence is enhanced in that environment up to a level $\delta B \approx B_0$ so as to ensure Bohm diffusion the maximum achievable energy turns out to be $E_{\rm max} \approx 10^4 GeV$ \cite{lc83b}, which is still an order of magnitude below the {\it knee}.

It is then clear that if the {\it knee} has to be reached, the level of magnetic turbulence in SNRs must be much higher than in the normal ISM and the turbulent magnetic field far exceed in strength the regular magnetic field. Evidence for such a phenomenon has indeed been found in recent years in the X-rays, thanks to the excellent data collected by the {\it Chandra} and {\it XMM} telescopes: these have shown that in several young Galactic SNRs non-thermal X-ray emission is concentrated in extremely thin rims at the outer edge of the remnant \cite{vink12}. The most obvious interpretation of the thickness of these filaments is as the distance high energy particles can travel before suffering severe energy losses. Assuming that the particle transport is governed by Bohm diffusion and losses are mostly due to synchrotron emission, one can estimate the thickness of the rims, which turns out to depend only on the magnetic field strength: $\Delta x=\sqrt{D_{\rm B}(E) \tau_{\rm sync}(E)}=0.04 B_{100}^{-3/2} {\rm pc}$, where $B_{100}$ is the local magnetic field strength in units of 100$\mu$G. Since the measured thickness is typically $\Delta x\approx 0.01{\rm pc}$, the implied magnetic fields are in the range a few $\times 100\mu$G. Similar estimates of the field result in other SNRs from analyses of fast time variability \cite{uchirxj}.

How these fields happen to be there has been a subject of much research and debates in the last few years (see \cite{bykovrev} for a thorough review). There are many different mechanisms for magnetic field amplification (MFA) in an environment such as that of a SNR: there are both fluid mechanisms, where the growth of an initial upstream perturbation is associated with fluid instabilities at the crossing of the shock (see e.g. \cite{giacajo}), and there are mechanisms associated to the presence of accelerated particles \cite{bell78,lc83b,ab06,bell04}. The main difference between these two broad categories is that the former only lead to MFA downstream of the shock, while the latter amplify the field in the upstream. As it is clear from Eqs.~\ref{eq:tacc} and \ref{eq:diffcoeff}, amplification in the downstream does not really solve the problem of decreasing the acceleration time and leading to a larger maximum energy: efficient scattering also in the upstream is required (first term in Eq.~\ref{eq:tacc}) unless the shock is almost perpendicular, in which case the particles quickly return to the shock in the upstream region, thanks to advection with the field lines. For more generic configurations enhanced scattering in the upstream is also needed to bring the maximum energy up to the {\it knee}. This requires the field to be amplified by some process associated with the presence of accelerated particles, which are the only messenger to run ahead of the shock and carry information about its presence.

The ability of fast particles at amplifying magnetic fields through the excitation of so-called streaming instabilities has been recognised since the '70s \cite{bell78}: the motion of relativistic particles through a plasma leads to the growth of resonant magnetic perturbations, if the associated streaming velocity is larger than the typical speed of hydromagnetic waves in a plasma, the Alfv\`en velocity, ${\rm v}_A=B_0/\sqrt{4 \pi n_i m_p}$, with $n_i$ the local density of ionised material and $m_p$ the proton mass. The mechanism behind the instability is the same responsible for particle scattering and isotropization: in a wave particle interaction there will be an exchange of momentum between the wave and the particle, which is maximised at resonance, namely when the wavelength of the wave is equal to the particle Larmor radius. This exchange of momentum tends to isotropise the accelerated particles in the wave frame, or equivalently to have them streaming at the Alfv\`en speed with respect to the background plasma.
While the resonant streaming instability likely plays a role in particle transport in the Galaxy, with CRs up to $\approx 200$ GeV being mainly scattered by the turbulence they generate themselves through this mechanism \cite{prl12}, in the context of a shock that is efficiently accelerating CRs the dominant phenomena are likely of a different nature. One important thing to remember is that SNRs must convert 10\% of the explosion kinetic energy into accelerate particles if they are the primary sources of CRs. This fact has two immediate consequences: 1)the density of accelerated particles in these systems is orders of magnitude larger than in the ISM and even more the associated current; 2)these particles take away from the system a non-negligible fraction of the energy, deeply modifying the structure of the shock. Both these effects modify the standard shock acceleration theory and need to be taken into account if one wants to make reliable predictions to quantitatively test how well the acceleration phenomenon is understood.

\begin{figure}
\includegraphics[scale=.5]{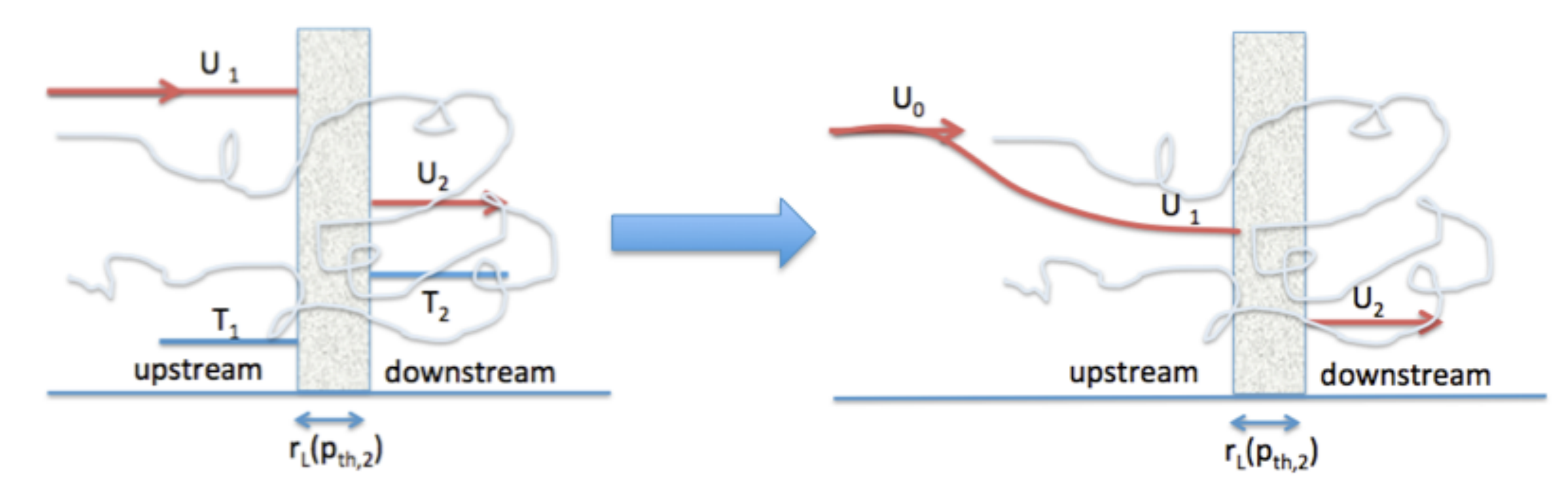}
\caption{Left panel: a shock that can be described in the test particle regime. Right panel: a shock that is accelerating particles efficiently. The red lines show the fluid velocity in the shock frame. On the right, one sees the formation of a precursor in the upstream, where the fluid is compressed and slowed down; then a subshock forms with a final velocity jump on the scale of the Larmor radius of the thermal particles downstream.}
\label{fig:modshock}
\end{figure}

Let us discuss the second effect first. This has been a subject of much work by several different groups employing different methods to build up the theory of "Non-linear diffusive shock Acceleration" (NLDSA hereafter, see \cite{amatorev,blasirev} and references therein). The acceleration mechanism implied is still the Fermi I process, but now a strong shock, that is efficiently accelerating particles, is no longer a sharp discontinuity in the plasma properties with a density increase (and a velocity decrease) by a factor of 4. The pressure carried by CRs in the upstream slows down the plasma ahead of the actual discontinuity leading to the formation of a shock precursor (see Fig.~\ref{fig:modshock}). The final jump in fluid velocity occurs then at a subshock with a compression ratio $R_s<4$. The overall shock compression ratio, however, between upstream infinity and downstream has become $R_T>4$: CRs have taken away energy from the shock, making it radiative. 

In terms of comparison between theory and data, total compression ratios $R_T>4$ have been deduced in at least 2 young SNRs, Tycho (right panel of Fig.~\ref{fig:CRspec}) and SN1006 \cite{warren05,gamil07}, from measurement of the distance between the contact discontinuity and the shock. Another hint of efficient acceleration is a lower downstream temperature than the Rankine-Hugoniot jump conditions would predict. This also has been inferred in a few SNRs \cite{rcw86,helder10,snr0509}.

The change in the shock dynamics has also important consequences on the spectrum of accelerated particles. In the test particle regime, Fermi mechanism leads to a particle distribution function that is a power-law with an index determined by the shock compression ratio. In the modified regime, no unique compression ratio can be defined: particles will experience a different compression ratio depending on how far in the upstream they diffuse. The highest energy particles, with diffusion path lengths equal to the extent of the precursor will experience the full $R_T$ and the spectral index will be $\gamma_p<4$ at the high energy end of the distribution. On the other hand, lower energy particles will sample progressively lower compression ratios, down to $R_s<4$, which implies $\gamma_p>4$. The end result is a concave spectrum, steeper than the test particle expectation at low energy and flatter at high energy. 

This is a very important prediction of this theory, but one for which we have contrasting evidence. On one hand there seems to be evidence for concave spectra in radio and X-ray observations of a few SNRs \cite{vink12}. But this emission is from electrons, and in addition some modelling of the history of the accelerators is involved, since radio emitting particles have synchrotron lifetimes comparable with the age of the SNR. On the other hand, wherever we observe $\gamma$-rays that are inferred to be of hadronic origin, the implied proton spectra are always steep \cite{damsteep}, both in the case of middle-aged SNRs interacting with clouds and accelerating particles up to relatively low energies (such as the already mentioned W44 and IC443), and in the case of young SNRs such as Cas A and Tycho \cite{fermicasa,fermitycho}, expected to host strong blast waves likely to be accelerating particles efficiently and up to energies close to the {\it knee}. 

A possible explanation for this apparent discrepancy between observations and the basic predictions of NLDSA might come from the first effect that was mentioned above, namely the phenomenon of MFA. The most likely interpretation for the amplified field inferred in a number of young SNRs is that they result from instabilities induced by the streaming of accelerated particles. The resonant streaming instability, important for Galactic particle transport, turns out to be somewhat suppressed in the presence of CR currents as large as those implied for a shock that is efficiently accelerating particles. However, while the growth rate of resonant Alfv\`en waves is lower in this context, a different, non resonant wave mode has been shown to grow very effectively \cite{bell04,ab09}: this is the so called "Bell's non resonant instability", that has received much attention in the last decade. The growth of this instability is very fast for young SNRs, and it can easily account for the large field values deduced from X-ray observations, for reasonable models of how the saturation occurs. However the field grows fast on very small scales, orders of magnitude smaller than the gyro-radius of the particles that carry the current. Such scales are irrelevant from the point of view of enhancing the particle scattering and decreasing the acceleration time, for which purpose resonant modes are needed. In fact, numerical experiments, performed with several different codes using different descriptions of the plasma (kinetic, hybrid, MHD), have shown that as the instability develops and the field strength grows, the typical scale of the turbulence also increases (see e.g. \cite{riquelme09} and references therein). Possible theoretical explanations behind this phenomenon involve a dynamo process seeded by the fact that only right hand polarised modes grow and hence the field has a net helicity \cite{bykovdyn}. 

Very interesting results are coming from Particle In Cell simulations of non relativistic shocks, which are finally becoming available in spite of the tremendous computational effort involved. These simulations, which solve the Maxwell's equations and the particles' equation of motion self consistently, are for the first time tracing the acceleration process from the very beginning and accounting for the effects of accelerated particles on the system \cite{damsim1,damsim2,damsim3}. The emerging picture is that MFA is initially induced from the very first particles that get accelerated and try to escape the system. During the acceleration process, the Larmor radius of these particles becomes so large that there is no turbulence to scatter them resonantly. Therefore they stream away from the shock, but in doing so they form a current that causes the growth of the Bell's instability. The dominant wavelength of the instability is initially very short but within few e-foldings a powerful inverse cascade develops that puts power at resonant wavelengths. The process is self-regulating \cite{schurebell}: the amplified field scatters the particles and limits their streaming, causing the current to decrease and the growth rate of perturbations with it; if this growth rate becomes too low, particles are again able to stream and to amplify the field. Also important is the fact that on average particle scattering is found to obey Bohm's law, with the diffusion coefficient computed in the amplified field: $D\approx cp/(e \delta B)$.  

This picture of the acceleration process at strong shocks brings with it several interesting consequences that help reconciling theory with observations. First of all the fact that the magnetic field is amplified to values much larger than its initial strength might help reconciling the theory of NLDSA with the observed steep spectra of young SNRs by effectively changing the compression ratio $R_T$ experienced by the particles. While one usually identifies $R_T$ with the fluid compression ratio, in reality, one should include the velocity of the scattering centers \cite{ziraptu,damsteep}, substituting $u_1$ ($u_2$) with $u_1+v_{A1}$ ($u_2+v_{A2}$), where $v_{A1}$ ($v_{A2}$) is the velocity of magnetic perturbations with respect of the fluid upstream (downstream) of the shock. Usually this change is irrelevant because $v_{A1}\ll u_1$ and $v_{A2}\ll u_2$. However, if the field is amplified to 100 times the value typical of the ISM, then the correction to the compression ratio experienced by the particles can account for steep spectra at shocks that are efficiently accelerating particles, as was shown in the case of Tycho. This young SNRs has been modelled within the framework of NLDSA with MFA and taking into account the effect of the amplified field in computing $R_T$: the result is that both its multiwavelength emission and spatial profile are well reproduced \cite{giodam12} if the SNR is converting about 15\% of its energy into CRs and accelerating protons up to 500 TeV (only a factor of a few short of the {\it knee}). 

Another consequence is actually a paradigm shift in terms of when the acceleration to the {\it knee} occurs. While this was thought to occur at few $\times$ 100 years after the SN explosion, more detailed calculations assuming Bell's mechanism of field amplification show that the {\it knee} is likely to be reached preferentially by remnants of core collapse SNe expanding in the dense slow wind that the progenitor star produced during its Red Supergiant phase. These SNRs enter the Sedov-Taylor phase in few tens of years, due to the high density of the medium in which they expand. Acceleration to the highest energies is then almost contemporary to the SN event \cite{cardillo15}, and it becomes very unlikely to see a PeVatron in action. So no wonder that we have not seen any yet.

\section{PWNe: the Pevatrons we have seen}
\label{sec:pwn}
The last statement is not completely correct. One should actually state that evidence is missing for a hadronic Pevatron, whereas we do see a leptonic Pevatron. In the Crab Nebula, one of the most famous and best studied objects in the sky, we do have direct evidence of PeV electrons and positrons. The Crab Nebula is the prototype of another class of extremely efficient Galactic accelerators: Pulsar Wind Nebulae. These objects are basically magnetised bubbles of relativistic electrons and positrons that shine through synchrotron emission and Inverse Compton scattering. Their ultimate source of energy is the rotational energy lost by a fast spinning magnetised neutron star, often observed as a pulsar. In the magnetosphere of such a star the conditions are such that efficient pair cascading takes place, and a large number of pairs (estimates are between $10^4$ and $10^7$ in the case of Crab) are produced by each electron that leaves the star surface: these objects are the primary antimatter factory in the Galaxy, and probably the main contributors to the ``rising positron fraction'' that has been observed by PAMELA \cite{pamela} and AMS02 \cite{ams02}.

As they leave the star magnetosphere, the pairs form a relativistic magnetised wind, that carries away most of the star spin down luminosity. In the case of young pulsars, such as Crab, this wind is confined by the surrounding SNR and hence has to slow down from its highly relativistic speed (corresponding to Lorentz factors in the range $10^4$-$10^6$) to the expansion velocity of the remnant, of order 1000 km/s. Such a transition cannot occur without dissipation and indeed a termination shock is established where the wind bulk energy is transformed into particle acceleration. 

The process of particle acceleration has several noticeable aspects in PWNe: 1) the acceleration efficiency is extremely high, with 20-30\% of the pulsar spin down power being converted into accelerated particles; 2) the maximum energy implied by observations is $\approx 1$ PeV; 3) the resulting particle spectrum is very flat at low energies (spectral index of the energy distribution between 1 and 1.5) and steepens at higher energies. In addition the acceleration process is associated with a shock that should not be accelerating particles at all according to the standard theory, namely a magnetised relativistic shock. Acceleration at such a shock cannot occur through the Fermi I process unless the magnetic field strength is such that $B^2/(4 \pi n m_e \Gamma c^2)<0.001$ \cite{sironi09} where $n$ is the pair plasma density, $m_e$ the electron mass and $\Gamma$ the wind Lorentz factor. Dynamical and radiation modelling of the Crab Nebula suggests that this condition is violated by most of the pulsar outflow, strongly disfavouring the Fermi I process as the main responsible of particle acceleration \cite{hepro4}. In any case, Fermi mechanism could only be at the origin of the steep spectrum high energy tail of the distribution, while most of the accelerated particles are in the flat low energy population.

Flat spectra are often seen in numerical experiments as a result of magnetic reconnection. The outcome of this process, however, depends on the context in which it occurs, and in the case of the Crab Nebula also this scenario has severe difficulties \cite{sironi11}, that are discussed together with other alternative proposals in another proceeding of this conference \cite{amatopwn15}. 

\section{Summary and Conclusions}
The last decade has been rich of developments in the quest for establishing what are the primary particle accelerators in the Galaxy. We have finally found direct evidence of hadronic acceleration in SNRs and we have built a theory that explains how the {\it knee} is reached and why it is difficult to see a hadronic PeVatron. On the other hand there are some aspects of the theory that are not fully satisfactory and at the same time not easy to improve. 
An example is the solution to the puzzle of steep SNR spectra that we discussed in Section~\ref{sec:nldsa}. The idea is that steep spectra and efficient acceleration go well together thanks to MFA and its effect on the velocity of the scattering enters: the viability of this process depends on details of how the turbulent cascade develops around a SNR shock that are currently unknown. It is to be hoped that progress on this subject will come from numerical studies, but currently the question cannot be considered as fully assessed. 

An even more critical issue is hidden in our current picture of PeVatrons: we identify these with SNRs expanding in the progenitor's wind, which is expected to have a toroidal magnetic field (hence perpendicular to the shock normal), and at the same time we assume that the acceleration occurs at a parallel shock (magnetic field parallel to the shock normal). One might think that Rayleigh-Taylor instabilities are likely to develop in the wind and make the field mostly parallel, but this is all very speculative. On the other hand it is intriguing that the only PeVatron we actually see, the Crab Nebula, is known to accelerate particles (leptons) with exceptionally high efficiency and to do it at a perpendicular shock. This fact might be taken to suggest that efficient acceleration processes, other than DSA, can be triggered by the presence of a shock, with an obvious example being magnetic reconnection. While DSA appears a well developed and mature theory, other potential acceleration mechanisms definitely deserve more attention.

\acknowledgments{Attendance of this conference has been made possible by the grant PRIN-INAF 2012}

\end{document}